\documentclass[aps,prl,reprint,superscriptaddress,twocolumn,showkeys,amsmath,amssymb]{revtex4-1}
\usepackage[english]{babel}
\usepackage[dvipsnames]{xcolor}
\usepackage{amsmath,amssymb,bbm,mathrsfs,bm,braket,graphicx,comment,amsfonts,dsfont,amsthm}
\usepackage[mathscr]{euscript}

\usepackage[T1]{fontenc}
\usepackage[latin9]{inputenc}
\usepackage{verbatim}

\usepackage{bbold}
\usepackage{tikz}
\usepackage{diagbox}
\usepackage[normalem]{ulem}
\usepackage{ifthen}

\makeatletter
\PassOptionsToPackage{caption=false}{subfig}

\usepackage{hyperref}
\hypersetup{
  colorlinks   = true, 
  urlcolor     = Blue, 
  linkcolor    = BrickRed, 
  citecolor   = PineGreen 
}

\IfFileExists{lmodern.sty}{\usepackage{lmodern}}{}

\renewcommand{\tilde}{\widetilde}

\renewcommand{\leq}{\leqslant}
\renewcommand{\geq}{\geqslant}

\newcommand{\Tr}{\operatorname{Tr}}

\newcommand{\calC}{\mathcal{C}}

\newcommand{\calO}{\mathcal{O}}

\makeatletter

\newcommand*{\wideboxed}[1]{\setlength{\fboxsep}{1ex}%
  \fbox{\m@th$\displaystyle#1$}}
\makeatother

\def\be{\begin{equation}}
\def\ee{\end{equation}}

\def\m{\mathcal}
\def\D{\Delta}

\def\sq3{1.7320508}

\newtheorem{prop}{Proposition}

\newtheorem{defn}{Definition}

\def\tcm{T.C.M. Group, Cavendish Laboratory, University of Cambridge, J.J. Thomson Avenue, Cambridge, CB3 0HE, UK}
\def\DAMTP{DAMTP, University of Cambridge, Wilberforce Road, Cambridge, CB3 0WA, UK}

\begin{document}
\setlength{\belowcaptionskip}{-10pt}

\title{Topologically ordered time crystals}

\author{Thorsten B. Wahl}
\affiliation{\DAMTP}
\author{Bo Han}
\affiliation{\tcm}
\author{Benjamin B\'eri}
\affiliation{\DAMTP}
\affiliation{\tcm}

\begin{abstract}
\vspace*{0.25em}
We define topological time crystals, a dynamical phase of periodically driven quantum many-body systems capturing the coexistence of intrinsic topological order with the spontaneous breaking of discrete time-translation symmetry. 
We show that many-body localization can stabilize this phase against generic perturbations and establish some of its key features and signatures, including a dynamical, time-crystal form of the perimeter law for topological order. 
We link topological and ordinary time crystals through three complementary perspectives: higher-form symmetries, quantum error-correcting codes, and a holographic correspondence. 
We also propose an experimental realization of a surface-code-based topological time crystal for the Google Sycamore processor. 
\vspace*{0.75em}
\end{abstract}

\maketitle

{\bf Introduction}---
Topologically ordered (TO) systems~\cite{Kitaev2003toriccode,Kitaev2006,*Levin-Wen,Nayak2008RMP}
display quantized nonlocal features such as fractional braiding statistics or groundstate degeneracy depending only on the topology of the underlying manifold.
These striking phenomena motivated numerous conceptual advances (e.g., symmetry-protected or  -enriched topological phases~\cite{Chen2010,Chen2013,HasanKane2010TIreview,Senthil2015,MesarosRan2013SET})
and applications [e.g., quantum error correcting (QEC) codes~\cite{Kitaev2003toriccode,dennisTopologicalQuantumMemory2002,FowlerMariantoniMartinisCleland2012surfacecode,TerhalRMP}] based on topological low-energy physics. 
More recently, studies of topological matter expanded beyond the low-energy regime to include far-from-equilibrium problems such as quenches~\cite{Caio2015,Hu2016,Wilson2016} or driven systems~
\cite{KitagawaBergRudnerDemler2010floquet,
RudnerLindnerBergLevin20132dfloquet,TitumBergRudnerRefaelLindner2016AFAI,
KhemaniLazaridesMoessnerSondhi2016TC,KeyserlingkSondhi2016floquetSPT,
*KeyserlingkSondhi2016floquet2,*RoyHarper2016FSPT1d,*PotterMorimotoVishwanath2016FSPT,Potirniche2017FSPTcoldatom,Flaschner2016FloquetBlochband}.

Time Crystals (TCs)\cite{Wilczek2012quantumTC,*ShapereWilczek2012classicalTC,*WatanabeOshikawa2015TC} are periodically driven (``Floquet'') systems that spontaneously break discrete time translation and a global internal symmetry~\cite{KhemaniLazaridesMoessnerSondhi2016TC,KeyserlingkSondhi2016floquetSPT,KeyserlingkKhemaniSondhi2016stability,Khemani2017,time_crystals_review,*Sacha2018,*Else2020}.
To form a dynamical phase of matter~\cite{Tauber2017,Oka2019}, TCs are protected from the drive-induced heating by many-body localization~\cite{Fleishman1980,gornyi2005interacting,*basko2006metal,NandkishoreHuse_review,*AltmanReview,*Abanin2017,*Alet2018,
ImbrieLIOMreview2017} (MBL): they are strongly disordered, with effective local degrees of freedom that remain inert if their energies are much below the driving frequency $2\pi /T$~\cite{abanin2015rigorous,KhemaniLazaridesMoessnerSondhi2016TC,KeyserlingkKhemaniSondhi2016stability,Khemani2017,time_crystals_review,Sacha2018,Else2020}.
TCs display spatiotemporal order~\cite{KeyserlingkKhemaniSondhi2016stability,Khemani2017,time_crystals_review,Sacha2018,Else2020}: 
local order parameters $O_j$ (at location $j$) exist such that $\langle \psi| O_j(mT) O_k(0)|\psi \rangle$ are nonconstant functions of the integer $m$ for large distances $|j-k|$ and for any eigenstate $|\psi\rangle$ of the time-evolution operator. 

In this work, we expand the TC concept to include TO. 
A key ingredient is to lift the global internal symmetries in TCs to ``higher-form'' global symmetries~\cite{Nussinov2009,*KapustinSeiberg2014TQFT,*GaiottoKapustinSeibergWillett2015GGS,*Yoshida2016,*Wen2019,*Lake2018,*Zhao2020}: 
TO can be viewed as the spontaneous breaking of these~\cite{Nussinov2009,KapustinSeiberg2014TQFT,GaiottoKapustinSeibergWillett2015GGS,Yoshida2016,Lake2018,Wen2019,Zhao2020}, and this provides a route for introducing topological time crystals (TTCs) along regular TC principles. 
While similar ideas were briefly mentioned earlier~\cite{KeyserlingkKhemaniSondhi2016stability,Khemani2017}, the considerations did not include MBL. 
Hence, even the existence of TTCs as dynamical phases is an open question. 
As TO requires two dimensions (2D) or higher, TTCs require MBL in 2D or above, where MBL may arise as a long-lived pre-thermal phase persisting beyond current experimental time-scales~\cite{chandran2016higherD,deRoeck2017Stability,Altman2018stability,Gopalakrishnan2019,Doggen20}. 
Focusing on this pre-thermal regime, we show that TTCs form a dynamical phase that is robust against perturbations,  and we establish some key TTC features and observable signatures, including a dynamical, time-crystal form of the perimeter law for TO~\cite{Wilson1974,Kogut1979,Polyakov2018,Huse2013LPQO,*2013Bauer_Nayak}.

To achieve our results, we formulate TTCs as Floquet-MBL versions of TO systems related to QEC codes.
On the technical level, this allows us to use a topological variant~\cite{WahlBeri2020TOMBL} of local integrals of motion (LIOMs)~\cite{serbyn2013local,*Huse_MBL_phenom_14,*chandran2015constructing,*ros2015integrals,*Inglis_PRL2016,*Rademaker2016LIOM,*Monthus2016,*Goihl2018,*Abi2017,*Abi2019}, an MBL framework that has been instrumental for ordinary TCs~\cite{KhemaniLazaridesMoessnerSondhi2016TC,KeyserlingkKhemaniSondhi2016stability,Khemani2017,time_crystals_review,Sacha2018,Else2020}.
Such topological LIOMs (tLIOMs) will give a similarly useful framework for TTCs. 
Conceptually, our approach highlights QEC codes as another unifying perspective bridging TCs and TTCs. 
We illustrate our findings, including both the QEC and the higher-form symmetry perspective, on TTCs in surface code systems~\cite{FowlerMariantoniMartinisCleland2012surfacecode}; 
this will allow us to highlight yet another unifying view: a holographic TTC--to--TC correspondence. 
We also explain how these  surface-code-based (pre-thermal) TTCs can be created in the Google Sycamore Processor~\cite{Arute2019googlesycamore,Google_SC} from native ingredients. 
In the Appendix, we provide a detailed discussion for the arguments in the main body of the paper and a comparison between regular TCs and TTCs.

{\bf Topological time crystals}--- 
We shall describe TTCs using the stabilizer formalism~\cite{Gottesman1997thesis,NielsenChuang2011QCQI}. 
We briefly summarize the ingredients for our purposes, focusing on a system with $N$ qubits, although the ideas are more general. 
The stabilizer group $\m{S}$ is an Abelian subgroup of the Pauli group; $\m{S}$ is generated by a set $\{A_P\}$ of suitable tensor products $A_P$ of Pauli operators (i.e., Pauli strings). 
Logical operators $\{W_\gamma\}$ are Pauli strings that commute with all $A_P$, but are not in $\m{S}$.
The $\{A_P\}$, together with a maximal mutually commuting subset of $\{W_\gamma\}$, form a complete mutually commuting set: their eigenstates uniquely specify a complete basis in the Hilbert space. 

We focus on lattice systems with nonchiral, Abelian TO, the kind of TO admitting a commuting-projector limit and MBL~\cite{Kitaev2006,Huse2013LPQO,*2013Bauer_Nayak,2015Potter,*Vasseur2016,WahlBeri2020TOMBL}. 
In the commuting-projector limit, the Hamiltonian is $H=\sum_P \alpha_P A_P$, with each $A_P$ local, i.e., with support  
(positions where $A_P$ differs from $\mathbb{1}$) 
of finite diameter much below the linear system size~\cite{Kitaev2003toriccode,Kitaev2006}. 
We will mostly work in 2D. 
There, the $\{W_\gamma\}$ run along noncontractible paths $\{\gamma\}$ and via their commutator algebra span  topologically degenerate eigenspaces~\cite{Kitaev2003toriccode,Kitaev2006}. 
MBL-like behavior is expected when $|\alpha_P|$ are large and random: 
in this case, adapting thermalization arguments~\cite{Gopalakrishnan2019} to TO~\cite{WahlBeri2020TOMBL} suggests that the thermalization timescale $t_\text{th}$ will be longer than exponential in the inverse perturbation strength away from the commuting-projector limit. 
All our statements below pertain to timescales below $t_\text{th}$. 
In this regime, one may accurately describe the system as one whose eigenstates are  simultaneous eigenstates of tLIOMs~\cite{WahlBeri2020TOMBL} $T_P = \tilde U^\dagger A_P \tilde U$ 
with support centered on that of $A_P$. 
Here, $\tilde U$ is a  local unitary~\cite{Bravyi2006,Chen2010,serbyn2013local,Huse_MBL_phenom_14,chandran2015constructing,ros2015integrals,Inglis_PRL2016,Rademaker2016LIOM,Monthus2016,Goihl2018,*Abi2017,Abi2019}, i.e., approximable to arbitrary accuracy by a constant-depth quantum circuit with 
gate length much below the linear system size~\cite{Wahl_PRX2017}.  
These tLIOMs imply TO in all eigenstates~\cite{WahlBeri2020TOMBL}. 
The dressed logical operator $\tilde{W}_\gamma = \tilde U^\dagger W_\gamma \tilde U$ contributes to the Hamiltonian only via couplings that decay exponentially with the length $|\gamma|$ of $\gamma$.
Unless stated otherwise, we assume large minimal $|\gamma|$ 
and ignore these exponentially small corrections (see Appendix~\ref{sec:prop1} for the general case).

\begin{figure}[t]
\includegraphics[width=0.33\textwidth]{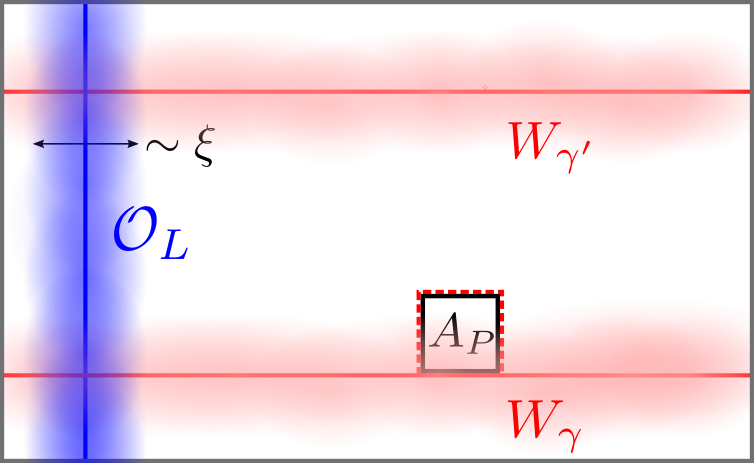}
\caption{Schematic of the support 
of bare logical operators $\mathcal{O}_L$, $W_\gamma$, and $W_{\gamma'}$ (solid lines) and of their dressed counterparts $\tilde{\mathcal{O}}_L$, $\tilde{W}_\gamma$, and $\tilde{W}_{\gamma'}$ (smeared lines of width set by the localization length $\xi$).
Multiplying $W_\gamma$ by a (bare) stabilizer $A_P$ (equivalently, $\tilde W_\gamma$ by the corresponding $T_P$) deforms its path $\gamma$ by $A_P$'s support (dashed). 
Via multiplying $W_\gamma$ by a suitable  $\prod_Q A_{Q}$, we get $W_{\gamma'}=W_\gamma \prod_Q A_{Q}$ along a path $\gamma'$ well separated from $\gamma$. 
}
\label{fig:deformation}
\end{figure}

As with ordinary TCs, we seek TTCs in Floquet MBL systems.
Over a driving period $T$, the time evolution is generated by the Floquet unitary $U_F \equiv \m{T} e^{-i \int^T_0 H(t) dt}$, where $\m{T}$ is the time-ordering operator and $H(t)$ is a local Hamiltonian (i.e., a sum of finite-range bounded-norm terms~\cite{Bravyi2006}) at time $t$. 
In particular, $\mathcal{O}(t+mT)=U_F^m \mathcal{O}(t) U_F^{m\dagger}$ for any operator $\m{O}$. 
With the ingredients above, we can define a TTC as follows.
\begin{defn}
\label{def:TTC}
A system is a TTC if 1)~the eigenstates $|\psi\rangle$ of $U_F$ are topologically ordered;    
2)~there exists a (dressed) logical operator $\tilde{W}_\gamma$ such that
$\langle \psi | \tilde{W}_\gamma (mT) \tilde{W}_\gamma(0) | \psi \rangle$ is a nonconstant function of the integer $m$
for any $|\psi\rangle$; 
3)~this property is robust to small, local, but otherwise generic perturbations to the drive.
\end{defn}
Compared with ordinary TCs, in 1) we replace spontaneous symmetry breaking in eigenstates~\cite{KeyserlingkKhemaniSondhi2016stability,Khemani2017,Sacha2018,time_crystals_review} by TO. 
Hence, in 2) instead of a local order parameter, we use a suitable nonlocal order parameter. 
($\tilde{W}_\gamma$ is, however, localized transverse to $\gamma$, cf. Fig.~\ref{fig:deformation}.)
Due to working with eigenstates $|\psi\rangle$, we have $\langle \psi | \tilde{W}_\gamma (mT) \tilde{W}_\gamma(0) | \psi \rangle=\pm\langle \psi | \tilde{W}_\gamma (mT) \tilde{W}_{\gamma^\prime}(0) | \psi \rangle$ for any  deformation $\gamma'$ of $\gamma$ via a stabilizer product (see Fig.~\ref{fig:deformation}). 
Hence, our definition also implies a generalization of long-range correlations for the nonlocal order parameter. 
We require 3) to capture a phase of matter: we exclude fine-tuned systems from our definition. 
We next consider the canonical drive structure that gives rise to a TTC. 
\begin{prop}
\label{prop:standard_form}
If an MBL Floquet unitary factorizes as $U_F = \tilde{\m{O}}_L e^{-i f(\{ T_P\})}$ with a dressed logical operator $\tilde{\m{O}}_L$ and an exponentially local function $f$ of tLIOMs  $T_P$, then such a factorization is robust and the system is a TTC. Here, exponentially local $f$ means that in
\begin{align}
f(\{T_P\}) = c_0 + \sum_P c_P T_P + \sum_{P,Q} c_{PQ} T_P T_Q + \ldots, \label{eq:tLIOMs} 
\end{align}
the $c_{PQR \ldots}$ decay exponentially with the largest distance between the centers of 
the supports of $T_Q$, $T_P$, $T_R$, $\ldots$.
\end{prop} 
The essence of the argument for this is as follows (see Appendix~\ref{sec:prop1} for details). 
We take $\tilde W_\gamma$ in Definition~\ref{def:TTC} to be conjugate to $\tilde{\m{O}}_L$, implying $\{\tilde{\m{O}}_L, \tilde W_\gamma\} = 0$. [Recall, $\tilde W_\gamma$ and $\tilde{\m{O}}_L$ are (smeared) logical Pauli operators.]
Then, the factorization states that $U_F$ is a local unitary~\cite{Bravyi2006,Chen2010,KeyserlingkKhemaniSondhi2016stability} and that $\theta_\gamma \equiv U_F \tilde W_\gamma U_F^\dagger \tilde W_\gamma=-\mathbb{1}$ (i.e., $U_F$ is odd in $\tilde{\m{O}}_L$). 
Such a factorization is robust if only $\theta_\gamma=\pm\mathbb{1}$ are possible: then no small perturbation can change $\theta_\gamma$.
To show $\theta_\gamma=\pm\mathbb{1}$, 
we deform $\tilde W_\gamma$  into  $\tilde W_{\gamma'}$ using a suitable $T_P$ product (cf. Fig.~\ref{fig:deformation}). 
For any $U_F(\tilde{\m{O}}_L,\tilde{W}_\gamma,\{T_P\})$ local unitary,
$\theta_{\gamma^{(\prime)}}$'s support, as $\tilde W_{\gamma^{(\prime)}}$'s, is localized around $\gamma^{(\prime)}$ in a width set by the localization length $\xi$. 
Yet, by $[T_P,\tilde{\m{O}}_L]=[T_P,\tilde{W}_\gamma]=0$, we have  $\theta_\gamma = \theta_{\gamma'}$, even if $\gamma$ and $\gamma'$ are much further apart than $\xi$.
This is possible only if $\theta_\gamma$ is a phase. 
Hence, $\theta_\gamma^2 = (\theta_\gamma \tilde W_\gamma)^2=(U_F \tilde W_\gamma U_F^\dagger)^2=\mathbb{1}$ which implies $\theta_\gamma = \pm\mathbb{1}$. 
The factorization incorporates eigenstate TO by construction and $\theta_\gamma=-\mathbb{1}$ implies robust period-$2T$ oscillations of the expectation values in Definition~\ref{def:TTC}: the system is a TTC. 

{\bf Comparison with other TCs}--- 
We first compare the TTC drive $U_F = \tilde{\m{O}}_L e^{-i f(\{ T_P\})}$ with the canonical drive structure of ordinary TCs~\cite{KeyserlingkKhemaniSondhi2016stability,Khemani2017,Sacha2018,time_crystals_review}:
There, instead of $\tilde{\m{O}}_L$ we have the (dressed) operator $\tilde{\m{P}}$ of the global symmetry, and for $f$ an MBL Hamiltonian ensuring spontaneous breaking of $\tilde{\m{P}}$ in all eigenstates. 
The TTC drive has similar ingredients in terms of higher-form symmetries~\cite{Nussinov2009,KapustinSeiberg2014TQFT,GaiottoKapustinSeibergWillett2015GGS,Yoshida2016,Lake2018,Wen2019,Zhao2020}: 
from this viewpoint, 2D TO in $f$ is the spontaneous breaking of ``$1$-form'' symmetries corresponding to (dressed) logical operators (such as $\tilde{\m{O}}_L$), while  the order parameters (``charged objects'') are their conjugates (such as $\tilde W_\gamma$). 
Owing to this unifying perspective, the TTCs we consider are arguably the most ``natural'' forms of TCs with intrinsic TO. 
They are distinct from TO drives that implement symmetry transformation of bulk anyons~\cite{PoFidkowskiVishwanathPotter2017chiralFloquet,*PotterMorimoto2017FSET,*Potter2018}, and by their intrinsic 2D TO also differ from 1D, symmetry-protected, or weak TTC phases~\cite{KeyserlingkSondhi2016floquetSPT,Giergiel_2019,*Chew2020,*bomantara2020twodimensional}. 

{\bf Comments and some generalizations}--- 
As in regular TCs, the decomposition of $U_F$ implies the organization of the TTC Floquet spectrum into eigenstate multiplets with rigid phase patterns~\cite{KhemaniLazaridesMoessnerSondhi2016TC,KeyserlingkKhemaniSondhi2016stability,Khemani2017,time_crystals_review}.
For $\tilde{\m{O}}_L^2=\mathbb{1}$, %
the $\tilde{\m{O}}_L$ prefactor imprints a robust
$\pi$ eigenphase difference.
This is exact for infinitely long $\tilde{\m{O}}_L$ even if its conjugate $\tilde{W}_\gamma$ has finite length; this is a TTC form of absolute stability~\cite{KeyserlingkKhemaniSondhi2016stability} (see Appendix~\ref{sec:TTCabsst} for details). 
This topological $\pi$ spectral pairing is in interplay with the topological degeneracy from TO: 
if
$f$ has $g$-fold spectral degeneracies [with degeneracy spaces labeled by $\tilde{\m{O}}_L$ and $\log_2(g)-1$ complementary dressed logical operators], the $\tilde{\m{O}}_L$ prefactor in $U_F$ imprints $\pi$ spectral pairing between $g/2$-fold degenerate eigenspaces. 
Thus far we focused on the case with logical operator algebra $W_\alpha W_{\beta} = \pm W_{\beta}W_\alpha$ and $W_\alpha^2=\mathbb{1}$; this corresponds to $\mathbb{Z}_2$ TO. 
More generally, one can consider, e.g.,  $W_\alpha W_{\beta} = e^{i 2\pi/n} W_{\beta}W_\alpha$ and $W_\alpha^n=\mathbb{1}$ as in  $\mathbb{Z}_n$ TO (corresponding to qudit QEC codes)~\cite{Kitaev2006}; in this case we get the TTC counterpart of ordinary $\mathbb{Z}_n$ TCs~\cite{KeyserlingkKhemaniSondhi2016stability,Khemani2017,time_crystals_review}.

{\bf Signatures of TTCs}--- While $U_F = \tilde \calO_L e^{-i f(\{T_P\})}$ implies a TTC in the sense of Definition~\ref{def:TTC}, the associated observable $\tilde W_\gamma$ depends on $\tilde U$, hence can be difficult to access experimentally. 
It is easier to access bare operators $W_\gamma$; we next focus on signatures in terms of these. 
In regular TCs one considers long-range correlations in eigenstates $|\psi\rangle$ (due to MBL, these have non-zero overlap with easily preparable product or stabilizer states)~\cite{KeyserlingkKhemaniSondhi2016stability,Khemani2017}. 
Analogously, we consider $\langle \psi | W_\gamma(mT) W_{\gamma^\prime}|\psi \rangle$,
where $\gamma^\prime$ is a deformation of $\gamma$ such that $\gamma$ and $\gamma^\prime$ are much further apart than 
$\xi$. 
We now argue that to get a TTC signal, the thermodynamic limit has to be qualified by how the lengths $|\gamma^{(\prime)}|$ (measured in units of lattice spacing) scale with system size. 
We consider a system with a single conjugate pair $\tilde O_L, \tilde W_\gamma$ of dressed logical operators (up to deformations).
As time crystallinity originates from the $\tilde W_\sigma$ (with $\sigma$ any deformation of $\gamma$) we must assess how these contribute to $W_{\gamma^{(\prime)}}$ (see also Appendix~\ref{sec:prop2}): 
Since each $\tilde W_\sigma$ is localized in a ribbon of width $\xi$ around $\sigma$ (cf. Fig.~\ref{fig:deformation}), the number of $\tilde W_\sigma$ for which $\gamma^{(\prime)}$ is within this ribbon (as required for $\tilde W_\sigma$ to contribute appreciably) is $\sim 2^{\xi |\gamma^{(\prime)}|}$.
In the expansion of $W_{\gamma^{(\prime)}}$ in the $\{T_P\},\tilde W_\gamma$ basis, these are however only the diagonal terms among $\sim 4^{\xi |\gamma^{(\prime)}|}$ terms; these are all roughly of equal weight because MBL provides no structure below the scale $\xi$. 
This implies that the $2T$-periodic signal decays as $\sim 2^{-\xi  (|\gamma|+|\gamma^{\prime}|)/2}$. 
While this may appear as a limitation, it is in fact a manifestation of a dynamical form of TO: 
it is the $2T$-periodic incarnation of the perimeter law for Wilson loop operators, familiar from the topological phase of lattice gauge theories~\cite{Wilson1974,*Kogut1979,*Polyakov2018}, including their MBL variants~\cite{Huse2013LPQO,*2013Bauer_Nayak}. 
The corresponding exponential decay with the \emph{perimeter} is to be contrasted to an exponential decay with the \emph{area} enclosed by the loops, which we would observe in a topologically trivial phase~\cite{Wilson1974,*Kogut1979,*Polyakov2018,Huse2013LPQO,*2013Bauer_Nayak}. 
Our result, thus, establishes a dynamical, time-crystal form of the perimeter law for TO. 
Overall, we find:
\begin{prop}
\label{prop:dressing}
For generic local perturbations, the time crystallinity of $U_F = \tilde \calO_L e^{-i f(\{T_P\})}$ is witnessed in the correlators of bare logical operators $W_\gamma$ only if the thermodynamic limit keeps the length of $\gamma$ finite. 
\end{prop} 

\noindent We emphasize, however, that finite $|\gamma|$ does not reduce the accuracy of the $2T$-periodicity, as follows from the TTC form of absolute stability we noted above and discuss in detail in Appendix~\ref{sec:TTCabsst}.

{\bf TTCs in surface codes}--- We next show how TTCs can arise in 2D surface codes.
We consider two examples: one motivated by bridging TCs and TTCs via a QEC perspective (which we shall explain), another based directly on the 1-form symmetry picture.
As we shall also explain, our examples illustrate a holographic TTC--to--TC correspondence, a dynamical version of the bulk-anyon-to-edge-symmetry relation of Refs.~\onlinecite{Severa_2002,freed2021topological,lichtman2020bulk,aasen2020topological}. 

For the QEC perspective, note that for 1D $\mathbb{Z}_2$ TCs $U_F = e^{-i H_0} e^{-iH_1}$ where $H_0=\sum_i g_i X_i$ and $H_1=\sum_i J_i Z_i Z_{i+1}$, and where $X_i$, $Z_i$ are Pauli operators on site $i$.
A TC arises for $g_i$ near $\pi/2$, i.e., $e^{-i H_0}$ near the $\mathbb{Z}_2 $ symmetry $\prod_i X_i$. 
The $\{Z_i Z_{i+1}\}$ and $\{X_i\}$ are complementary sets of stabilizer generators: respectively for a repetition code~\cite{bomantara2021quantum}
(related to $\mathbb{Z}_2$-symmetry-breaking states), 
and for product states; the generators of one code flip those of the other. 
(Viewing the system as a fermion chain via Jordan-Wigner transformation, it also exemplifies a Majorana TTC with 1D TO~\cite{KhemaniLazaridesMoessnerSondhi2016TC,KeyserlingkSondhi2016floquetSPT}.)

\begin{figure}[t]
\begin{picture}(220,160)
\put(0,0){
\begin{tikzpicture}[scale=0.6]

 \draw (0,0)--(1,0)--(2,0)--(3,0)--(4,0);
 \draw (0,1)--(1,1)--(2,1)--(3,1)--(4,1);
 \draw (0,2)--(1,2)--(2,2)--(3,2)--(4,2);
 \draw (0,3)--(1,3)--(2,3)--(3,3)--(4,3);
 \draw (0,4)--(1,4)--(2,4)--(3,4)--(4,4);
 
 \draw (0,0)--(0,1)--(0,2)--(0,3)--(0,4);
 \draw (1,0)--(1,1)--(1,2)--(1,3)--(1,4);
 \draw (2,0)--(2,1)--(2,2)--(2,3)--(2,4);
 \draw (3,0)--(3,1)--(3,2)--(3,3)--(3,4);
 \draw (4,0)--(4,1)--(4,2)--(4,3)--(4,4);

 \draw (0,0) arc (270:90:0.5cm);
 \draw (0,2) arc (270:90:0.5cm);
 \draw (0,4) arc (180:0:0.5cm);
 \draw (2,4) arc (180:0:0.5cm);
 \draw (1,0) arc (180:360:0.5cm);
 \draw (3,0) arc (180:360:0.5cm);
 \draw (4,1) arc (-90:90:0.5);
 \draw (4,3) arc (-90:90:0.5);
 
 \node at (0.5,3.5){$Z$};
 \node at (2.5,3.5){$Z$};
 \node at (4.25,3.5){$Z$};
 
 \node at (1.5,2.5){$Z$};
 \node at (3.5,2.5){$Z$};
 \node at (-0.25,2.5){$Z$};
 
 \node at (0.5,1.5){$Z$};
 \node at (2.5,1.5){$Z$};
 \node at (4.25,1.5){$Z$};
 
 \node at (1.5,0.5){$Z$};
 \node at (3.5,0.5){$Z$};
 \node at (-0.25,0.5){$Z$};

 \node at (0.5,0.5){$X$};
 \node at (2.5,0.5){$X$};
 \node at (1.5,-0.25){$X$};
 \node at (3.5,-0.25){$X$};
 \node at (1.5,1.5){$X$};
 \node at (3.5,1.5){$X$};
 \node at (0.5,2.5){$X$};
 \node at (2.5,2.5){$X$};
 \node at (1.5,3.5){$X$};
 \node at (3.5,3.5){$X$};
 \node at (0.5,4.25){$X$};
 \node at (2.5,4.25){$X$};
 
 \begin{scope}[shift={(-6,-6.6)}]
 \draw (6,3)--(7,3)--(7,4)--(6,4)--cycle;
 \node at (6.5,3.5){$Q$};
 \node at (7.5,3.5){$=$};
 \draw (8,3)--(9,3)--(9,4)--(8,4)--cycle;
 \node at (8,3){$Q$};
 \node at (9,3){$Q$};
 \node at (9,4){$Q$};
 \node at (8,4){$Q$};
 \end{scope}
 
 \begin{scope}[shift={(-2,-4.3)}]
 \draw (6,1)--(7,1);
 \draw (6,1) arc (180:0:0.5cm);
 \node at (6.5,1.25){$Q$};
 \node at (7.5,1.25){$=$};
 \draw (8,1)--(9,1);
 \draw (8,1) arc (180:0:0.5);
 \node at (8,1){$Q$};
 \node at (9,1){$Q$};
 \end{scope}
 
 \begin{scope}[shift={(3,-4.4)}]
 \node at (6.5,1.25){$Q=X,Z$};
 \end{scope}
 
 \draw [line width=0.5mm,red] (0,1)--(4,1);
 \node [right] at (3.85,0.7){$Z_L$};
 \draw [line width=0.5mm, blue] (1,0)--(1,4);
 \node [above left] at (1.9,3.9){$X_L$};

\begin{scope}[shift={(3.2,-0.7)},scale=0.78]

\foreach \y in {-1,0,...,6}{
\ifthenelse{\y=2 \OR \y=3}{\draw (3,\y)--(5,\y); \draw (8,\y)--(10,\y)}{ \draw (3,\y)--(10,\y)};
    }
 
\foreach \x in {3,4,...,10}{
\ifthenelse{\x=6 \OR \x=7}{\draw (\x,-1)--(\x,1); \draw (\x,4)--(\x,6)}{ \draw (\x,-1)--(\x,6)};
    }
 
\foreach \x in {-1,1,3,5}{
 \draw (3,\x) arc (270:90:0.5cm);
  \draw (10,\x) arc (-90:90:0.5);
  \node at (2.8,\x+0.5){$X$};
   \node at (10.25,\x+0.5){$X$};
 }

\foreach \x in {3,5,7,9}{
 \draw (\x,6) arc (180:0:0.5);
  \draw (\x,-1) arc (180:360:0.5cm);
  \node at (\x+0.5,6.23){$X$};
  \node at (\x+0.5,-1.24){$X$};
}

\foreach \x in {3,5,...,9}{
\foreach \y in {-1,1,...,5}{
\ifthenelse{\x=5 \OR \x=6 \OR \x=7}{\ifthenelse{\y=1 \OR \y=2 \OR \y = 3}{}{\node at (\x+0.5,\y+0.5){$Z$}}}{\node at (\x+0.5,\y+0.5){$Z$}};
    }
    }

\foreach \x in {4,6,...,8}{
\foreach \y in {0,2,...,4}{
\ifthenelse{\x=5 \OR \x=6 \OR \x=7}{\ifthenelse{\y=1 \OR \y=2 \OR \y = 3}{}{\node at (\x+0.5,\y+0.5){$Z$}}}{\node at (\x+0.5,\y+0.5){$Z$}};
    }
    }

\foreach \x in {3,5,...,9}{
\foreach \y in {0,2,...,4}{
\ifthenelse{\x=5 \OR \x=6 \OR \x=7}{\ifthenelse{\y=1 \OR \y=2 \OR \y = 3}{}{\node at (\x+0.5,\y+0.5){$X$}}}{\node at (\x+0.5,\y+0.5){$X$}};
    }
    }

\foreach \x in {4,6,...,8}{
\foreach \y in {-1,1,...,5}{
\ifthenelse{\x=5 \OR \x=6 \OR \x=7}{\ifthenelse{\y=1 \OR \y=2 \OR \y = 3}{}{\node at (\x+0.5,\y+0.5){$X$}}}{\node at (\x+0.5,\y+0.5){$X$}};
    }
    }
 
\draw[line width=0.5mm, red] (5,1)--(5,4)--(8,4)--(8,1)--cycle;
\draw [line width=0.5mm, blue] (6,4)--(6,6);

\draw (6,1) arc (180:0:0.5);
\draw (6,4) arc (180:360:0.5cm);
 \draw (8,2) arc (270:90:0.5cm);
  \draw (5,2) arc (-90:90:0.5);
  \node at (6.5,1.25){$X$};
  \node at (6.5,3.75){$X$};
  \node at (7.75,2.5){$X$};
  \node at (5.25,2.5){$X$};

\node at (6.5,6.27){$X_L$};
\node at (5.5,3.6){$Z_L$};

\end{scope}
 
\end{tikzpicture}
}
\put(0,150){\textbf{a}}
\put(110,150){\textbf{b}}
\put(0,25){\textbf{c}}
\end{picture}
\caption{a: Odd-by-odd surface code. 
b: Even-by-even surface code with a hole. c: 
The stabilizer generators $A_P=\prod_{i\in P} Q_i$ (with $Q = X, Z$, depending on the plaquette $P$).
The logical operators $Q_L=\prod_{i \in Q_L} Q_i$ (red for $Q=Z$, blue for $Q=X$) are Pauli strings along noncontractible paths, deformable via suitable products of $A_P$.
} 
\label{fig:surfacecode}
\end{figure} 

The surface code counterpart of this is $U_F = e^{-i H_0} e^{-iH_1}$ with $H_{1} = -\sum_{P} J_P  A_P$ comprised of surface code stabilizer generators $A_P=\prod_{i\in P} Q_i$ (with $Q = X, Z$, depending on the plaquette $P$, see Fig.~\ref{fig:surfacecode}), and again $H_0 = \sum_i g_i X_i$. 
We first focus on an $N=d^2$ system with $d$ odd [Fig.~\ref{fig:surfacecode}(a)]. 
As in regular TCs, we inspect $g_i=\pi/2$; 
in this case $U_F = X_L e^{-i H_{1}'}$ with $X_L$ a logical operator and $e^{-i H_{1}'}$ is obtained by absorbing the $A_P$ product $X_Le^{-i H_0(g_i=\pi/2)}$ into $e^{-i H_{1}}$.
This $U_F$ has the canonical structure  and hence the associated robustness. 
Does the construction generalize to other layouts, topologies, or other codes for $H_1$?
This is guaranteed if (i) the $A_s$ are purely $X$- or $Z$-strings (they generate a Calderbank-Shor-Steane code~\cite{NielsenChuang2011QCQI}),
(ii) the $Z$-stabilizers are even-length, and (iii) the code has an odd-length $Z_L=\prod_{i \in Z_L} Z_i$:
then $U_0=e^{-iH_0(g_i=\pi/2)}\propto \prod_i X_i$ satisfies $[U_0,A_s]=\{U_0,Z_L\}=0$; hence,  
$U_0$ is a logical operator odd in $X_L$.
(For similar conditions in the QEC context, cf. Ref.~\onlinecite{Venn2020}.)

For the bare $Z_L$ to give TTC signatures when $N\rightarrow\infty$ in this layout, one may require that perturbations respect $\tilde Z_L=Z_L$ thus sidestepping Proposition~\ref{prop:dressing}. 
Alternatively, one may allow generic perturbations, but change the aspect ratio with $N$ so that $Z_L$ has fixed length; this leads to a quasi-1D surface code when $N\rightarrow \infty$.

Maintaining a 2D $N\rightarrow\infty$ limit while retaining TTC signatures in bare operators is possible, e.g., in a surface code with a hole [Fig.~\ref{fig:surfacecode}(b)]. 
For $U_F$, here we adopt the 1-form symmetry viewpoint: we directly get $U_F (g_i=\pi/2) \propto X_L e^{-i H_{1}}$ by taking $H_0 = \sum_{i \in \calC} g_i X_i$, where $\calC$ is any path from the hole to the outer boundary.
[In this way, we can relax conditions (i-iii) above, but we give up having a purely 2D $H_0$.]
The TTC signatures can survive in bare $Z_L$ operators, even with generic perturbations, provided the $N\rightarrow\infty$ limit keeps the hole perimeter fixed.  
There is still a tradeoff in separating $\gamma$ from $\gamma^\prime$ (both encircling the hole) beyond $\xi$ while keeping $|\gamma^{(\prime)}|\xi$ finite for an appreciable signal;
the most favorable regime to observe TTCs is that of small $\xi$ (strong MBL).

{\bf Holographic TTC--to--TC correspondence---} 
The systems in Fig.~\ref{fig:surfacecode} also exemplify a dynamical variant of TO with gapped boundaries: they are TTCs (and as such TO MBL) with MBL boundaries. 
The phases of 1D (clean) systems with global symmetries are equivalent to the boundary phases of (clean) nonchiral 2D TO systems; e.g., the 1-form symmetry $X_L$ implements a $\mathbb{Z}_2$ symmetry for a boundary on which $Z_L$ can end, $Z_L$ implements the boundary order parameter, and anyonic symmetries imply boundary dualities~\cite{Severa_2002,freed2021topological,lichtman2020bulk,aasen2020topological}.
These relations naturally generalize to a link between 1D regular TCs and 2D TTCs. 
In particular, for $U_F = \tilde{\m{O}}_L e^{-i f(\{ T_P\})}$, if a boundary $B$ along $O_L$ exists, such that $\tilde{W}_\gamma$ can end on $B$ while still commuting with all tLIOMs and anticommuting with $\tilde O_L$ (such $B$ is one of the boundaries parallel to the blue string $O_L$ in Fig.~\ref{fig:surfacecode}a), one can view $B$ as a regular TC with $\mathbb{Z}_2$ symmetry via $\tilde{\m{O}}_L$. In contrast, with $B$ located as before but with boundary tLIOMs that anticommute with $\tilde{W}_\gamma$ terminating on $B$, one can view $B$ as an MBL paramagnet with the same symmetry. 
It would be interesting to explore how this might generalize and enrich (e.g., via correspondence to 1D symmetry-protected TCs~\cite{KeyserlingkSondhi2016floquetSPT,time_crystals_review}) the possible TTC phases or the kind of insights it might provide into 1D TC phase diagrams.

\begin{figure}[t]
\begin{picture}(200,160)
\put(-20,25){\centerline{\includegraphics[width=0.27\textwidth]{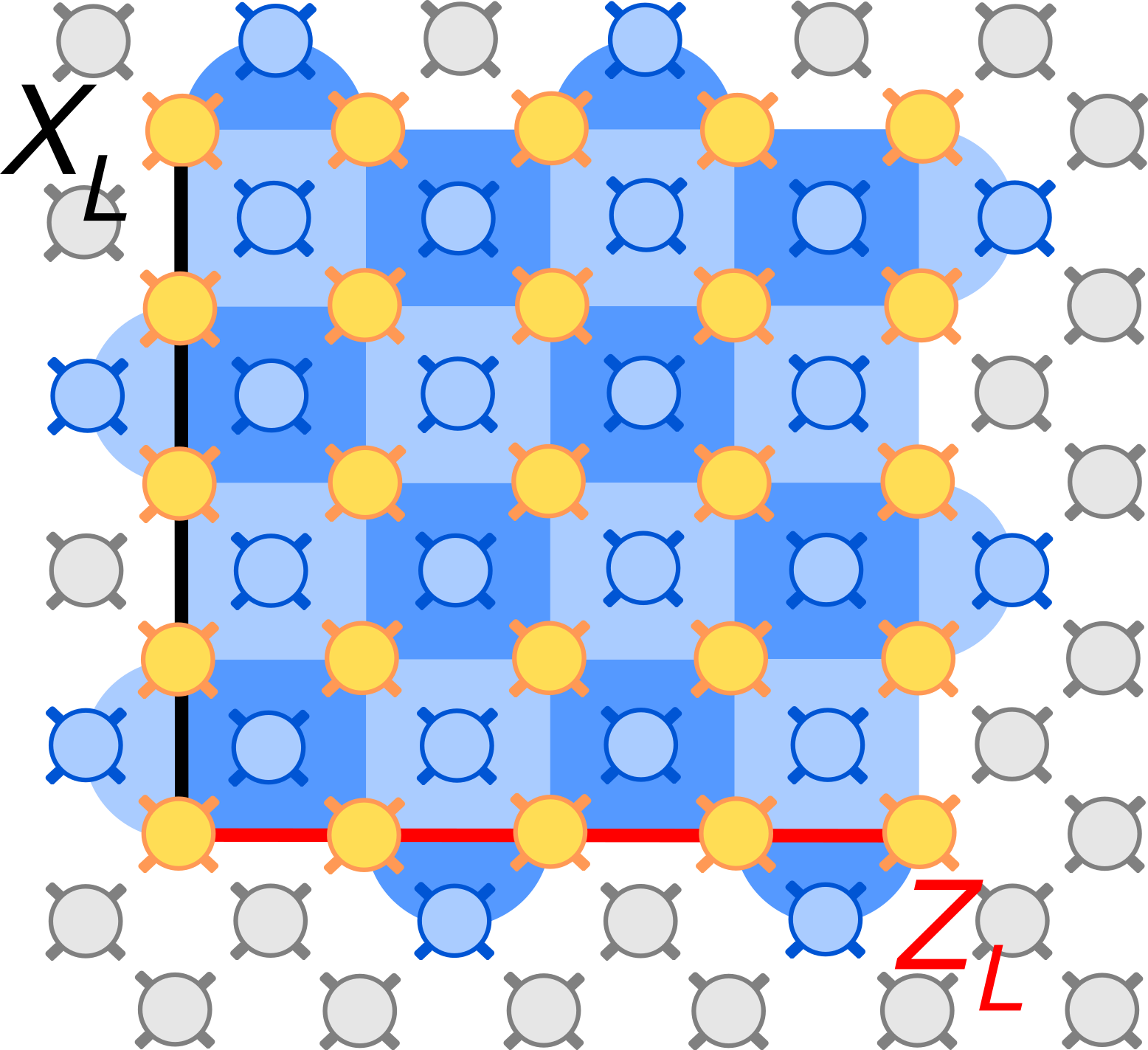}}}
\put(-10,150){\textbf{a}}
\end{picture}
 \vspace{10pt} 
 \\
\begin{picture}(200,80)
\put(0,0){

\begin{tikzpicture}[scale=0.7]
 \draw (0,0)--(1,0)--(1,1)--(0,1)--cycle;
 \node at (1.5,0.5){$=$};
 \node at (0.5,0.5){$U_{Z_P}$};
 
 \draw (2,2)--(9,2);
 \draw (2,1.5)--(9,1.5);
 \draw (2,1)--(9,1);
 \draw (2,0.5)--(9,0.5);
 \draw (2,-0.5)--(4.5,-0.5);
 \draw (6,-0.5)--(9,-0.5);
 \node at (5.25,-0.5){\small{$e^{-i \D t Z}$}};
 
 \draw [fill,black] (2.5,2) circle (2pt);
 \draw (2.5,-0.5) circle (5pt);
 \draw (2.5,2)--(2.5,-0.68);
 
 \draw [fill,black] (2.9,1.5) circle (2pt);
 \draw (2.9,-0.5) circle (5pt);
 \draw (2.9,1.5)--(2.9,-0.68);
 
 \draw [fill,black] (3.3,1) circle (2pt);
 \draw (3.3,-0.5) circle (5pt);
 \draw (3.3,1)--(3.3,-0.68);
 
 \draw [fill,black] (3.7,0.5) circle (2pt);
 \draw (3.7,-0.5) circle (5pt);
 \draw (3.7,0.5)--(3.7,-0.68);
 
 \draw (4.5,0)--(4.5,-1)--(6,-1)--(6,0)--cycle;
 
 \draw [fill,black] (6.8,0.5) circle (2pt);
 \draw (6.8,-0.5) circle (5pt);
 \draw (6.8,0.5)--(6.8,-0.68);
 
 \draw [fill,black] (7.2,1) circle (2pt);
 \draw (7.2,-0.5) circle (5pt);
 \draw (7.2,1)--(7.2,-0.68);
 
 \draw [fill,black] (7.6,1.5) circle (2pt);
 \draw (7.6,-0.5) circle (5pt);
 \draw (7.6,1.5)--(7.6,-0.68);
 
 \draw [fill,black] (8,2) circle (2pt);
 \draw (8,-0.5) circle (5pt);
 \draw (8,2)--(8,-0.68);
 
 \node at (1.7,-0.5){$|0\rangle$};
  \node at (9.4,-0.5){$|0\rangle$};
 
 \draw (0,-2)--(0,-3)--(1,-3)--(1,-2)--cycle;
 \node at (0.5,-2.5){$U_{X_P}$};
 \node at (1.5,-2.5){$=$};
 \draw (2.25,-2.5)--(3,-2.5);
 \draw (3,-2)--(3,-3)--(4.25,-3)--(4.25,-2)--cycle;
 \draw (4.25,-2.5)--(5.25,-2.5);
 \draw (5.25,-2)--(5.25,-3)--(6.25,-3)--(6.25,-2)--cycle;
 \draw (6.25,-2.5)--(7.25,-2.5);
 \draw (7.25,-2)--(7.25,-3)--(8.7,-3)--(8.7,-2)--cycle;
 \draw (8.7,-2.5)--(9.45,-2.5);
 
 \node at (5.75,-2.5){$U_{Z_P}$};
 \node at (3.625,-2.5){$\sqrt{Y}^{\otimes 4}$};
 \node at (8,-2.5){$\sqrt{Y^\dagger}^{\otimes 4}$};
 
\end{tikzpicture}
}
\put(-10,100){\textbf{b}}
\end{picture}
\caption{a: The Google Sycamore device of Ref.~\onlinecite{Google_SC}, with $N=25$ data qubits (gold), realizes the system of Fig.~\ref{fig:surfacecode}a.
(Measure qubits are in blue; the figure is redrawn from Ref.~\onlinecite{Google_SC}.)
Dark (light) blue shaded areas mark $X$ ($Z$) stabilizers.
b:~The plaquette evolution $U_{Z_P}=\exp(-i \Delta t \prod_{j \in P} Z_j)$ on four data qubits. 
The surrounded measure qubit starts in $|0\rangle$; the CNOTs couple to the data qubits.
The $X$-plaquette evolution arises via conjugating by $\sqrt{Y}$ on the data qubits.}
\label{fig:sycamore}
\end{figure}

{\bf TTC in Google Sycamore}--- The surface code groundstate, anyons, and logical operators have seen recent Google Sycamore realizations~\cite{Satzinger2021,Google_SC} and the same platform has been argued to be excellently suited for realizing TCs~\cite{ippoliti2021manybody}. (See also Ref.~\onlinecite{frey2021simulating} for an IBM realization.)
We now describe how the Sycamore can be used to create and detect a TTC. 
As in Ref.~\onlinecite{FowlerMariantoniMartinisCleland2012surfacecode}, we divide the square grid of qubits into ``data qubits'' and ``measure qubits'' (Fig.~\ref{fig:sycamore}a). 
Data qubits are to be evolved under $U_F$; measure qubits facilitate the desired multi-qubit gates. 
For the plaquette evolution $\exp({-i \Delta t \prod_{j \in P} Z_j})$, this follows the standard approach (Fig.~\ref{fig:sycamore}b)~\cite{NielsenChuang2011QCQI}. 
For $\exp({-i \Delta t \prod_{j \in P} X_j})$, one conjugates the above by 
$\sqrt{Y}$
on data qubits, using $\sqrt{Y} Z \sqrt{Y^\dagger} = X$.
For $\exp{(-iH_0)}$, one applies phase gates on data qubits. 
This completes the creation of $U_F$ for a surface code TTC.
All the ingredients, namely the phase gate, $\sqrt{Y}$, and CNOT (via conjugating CZ by $\sqrt{Y}$) are natively available in Sycamore~\cite{Arute2019googlesycamore}. 
Detecting time crystallinity can proceed via the interferometric protocol demonstrated in Ref.~\onlinecite{Satzinger2021}.
The robustness of TTCs guarantees resilience against static gate inaccuracies.
The levels of noise (e.g., fluctuating gate parameters, decoherence) have been estimated to be compatible with TCs~\cite{ippoliti2021manybody} and we expect the same for TTCs.

{\bf Conclusions}---  We have defined TTCs and showed that, combined with MBL, they form a (pre-thermal) dynamical phase.
Higher-form symmetries and QEC codes offer complementary ways to link regular TCs to TTCs,  
while the holographic correspondence between TTCs and their MBL boundaries, as illustrated by our surface code example, offers a reverse link to regular TCs. 
Logical operators serve both as (emergent) symmetries and as order parameters for TTCs and this leads to interesting interplay, not only between spectral pairing patterns and topological degeneracies, but also between MBL and the nonlocality of these operators. 
The latter interplay results in a dynamical, time-crystal form of the perimeter law. 
A practical implication of this fundamental result is that one must qualify the thermodynamic limit (as, e.g., in a surface code with a hole) to maintain the observability of TTCs via bare operators. 

For realizing TTCs, the most favorable settings are those with short localization lengths (strong MBL); then TTCs can appear already in moderate-sized systems.  
In particular, the Google Sycamore~\cite{Arute2019googlesycamore}, including its version used in recent QEC experiments~\cite{Google_SC}, has all the ingredients for creating such a TTC.

In the future, it would be interesting to explore  TTCs in other TO QEC code systems (including those above 2D), such as Floquet-MBL versions of color codes~\cite{Bombin2DCC,*Bombin3DCC}, where one could study the role of the transversal gates, or TTCs in fracton systems~\cite{Haah2011,*X-cube,*Fractons}.

{\bf Acknowledgments}---
This project was supported by the ERC Starting Grant No.~678795 TopInSy and the EPSRC grant EP/V062654/1. TBW acknowledges support through the Royal Society Research Fellows Enhanced Research Expenses 2021 RF\textbackslash{ERE}\textbackslash{210299}.

\appendix

\section{Demonstration of Proposition 1}
\label{sec:prop1}

We first state Proposition 1 in a form 
assuming
that the length $|L|$ of $\tilde{\m{O}}_L$'s path $L$ far exceeds the localization length $\xi$ (allowing us to take $|L|\rightarrow \infty$), but allows  a finite 
$|\gamma|$ for $\tilde{W}_\gamma$. 
(Proposition 1 in the main text assumes $|L|,|\gamma|\rightarrow \infty$.)
Here, and in Proposition 1, we use a convention where $L$ and $\gamma$ are the shortest among their respective deformations (cf. Fig. 1 of the main text). 
Below, the diameter of a dressed operator $\tilde{O}=\tilde{U}^\dagger O \tilde{U}$ means the diameter of its bare counterpart's, i.e., $O$'s, support.
(An operator's support is the set of positions where it does not act as $\propto\mathbb{1}$.)

\begin{prop}
\label{prop:standard_form}
If an MBL Floquet unitary factorizes as $U_F = \tilde{\m{O}}_L e^{-i f(\{ T_P\},\tilde{W}_\gamma)}$ in terms of the operators $\tilde{\m{O}}_L$, $\{T_P\}$, $\tilde{W}_\gamma$, and function $f$, all described below, then such a factorization is robust and the system is a TTC.
Here, $\{T_P\}$ is the set of tLIOMs, $\tilde{\m{O}}_L$ and $\tilde{W}_\gamma$ are corresponding mutually conjugate dressed logical operators (hence $[\tilde{\m{O}}_L,T_P] = [\tilde{W}_\gamma,T_P]=\{\tilde{W}_\gamma,\tilde{\m{O}}_L\}=0$) along a path $L$ and a possibly finite-length path $\gamma$, respectively,
and 
\begin{align}
&f(\{T_P\},\tilde{W}_\gamma) = \left(c_0 + \sum_P c_P T_P + \sum_{P,Q} c_{PQ} T_P T_Q + \ldots\right) \notag \\
&+\tilde{W}_\gamma\left(c^\prime_0 + \sum_P c^\prime_P T_P + \sum_{P,Q} c^\prime_{PQ} T_P T_Q + \ldots\right),\label{eq:tLIOMs} 
\end{align}
where the coefficients $c_{PQ}$, $c_{PQR}$, $c_{PQRS}$, $\ldots$ decay exponentially with the diameter of the corresponding tLIOM products, and $c^\prime_{0}$, $c^\prime_{P}$, $c^\prime_{PQ}$, $\ldots$ decay exponentially with the diameter of $\tilde{W}_\gamma$, $\tilde{W}_\gamma T_P$, $\tilde{W}_\gamma T_P T_Q$, $\ldots$, respectively. 
\end{prop} 

We demonstrate this using an approach similar to that in Ref.~\onlinecite{KeyserlingkKhemaniSondhi2016stability}. 
Since $\tilde{\m{O}}_L$ is a (dressed) logical Pauli operator we have $\tilde{\m{O}}^2_L=\mathbb{1}$.
Therefore the factorization $U_F = \tilde{\m{O}}_L e^{-i f(\{ T_P\},\tilde{W}_\gamma)}$ is the statement that $U_F$ is odd in $\tilde{\m{O}}_L$, thus it anticommutes with $\tilde{W}_\gamma$. 
The dependence $f(\{ T_P\},\tilde{W}_\gamma)$ is implied by TO MBL~\cite{serbyn2013local,Huse_MBL_phenom_14,chandran2015constructing,ros2015integrals,Inglis_PRL2016,Rademaker2016LIOM,Monthus2016,Goihl2018,*Abi2017,Abi2019,WahlBeri2020TOMBL} and the decay of $c^{(\prime)}$ follows both from TO MBL and from $U_F$ being a finite-time evolution by a local Hamiltonian and hence~\cite{Bravyi2006,Chen2010} a local unitary, i.e., approximable to arbitrary accuracy by a constant-depth quantum circuit with much smaller gate length than the linear system size.

Our first goal is to show that this factorization is robust to perturbations. 
To this end, 
allowing for any local unitary $U_F(\tilde{\m{O}}_L,\tilde{W}_\gamma,\{T_P\})$, 
we consider $\theta_\gamma =  U_F {\tilde W}_\gamma U_F^\dagger {\tilde W}_\gamma$ which indicates to what extent ${\tilde W}_\gamma$ and $U_F$ (anti)commute. 
As ${\tilde W}_\gamma$ is a dressed logical operator, its support 
is localized in a ribbon of width $\xi$ around $\gamma$ (to exponential accuracy, i.e., with tails that decay exponentially in the direction transverse 
to $\gamma$ on a scale set by $\xi$). 
Since $U_F$ is a local unitary, the support of $\theta_\gamma$ is also similarly localized around $\gamma$.

\begin{figure}[t!]
\includegraphics[width=0.33\textwidth]{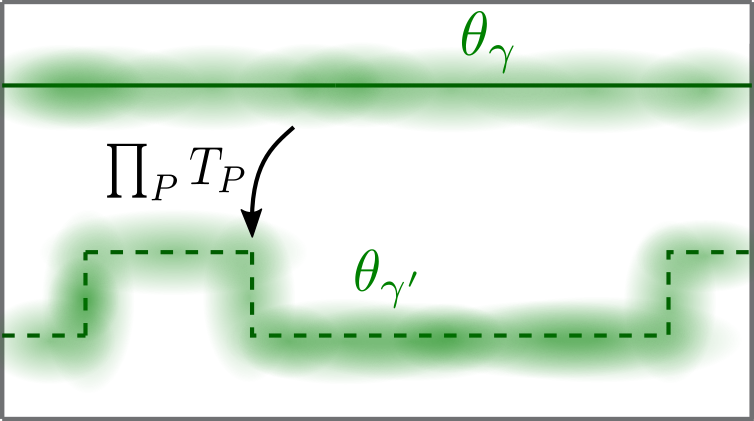}
\caption{The supports of $\theta_\gamma$ and $\theta_{\gamma'}$, illustrated as smeared lines around the paths $\gamma$ (solid line) and $\gamma'$ (dashed line), respectively. The relation $\theta_\gamma = \theta_{\gamma'}$ that we establish implies $\theta_\gamma \propto \mathbb{1}$, which in turn is shown to imply $\theta_\gamma =\pm \mathbb{1}$.}
\label{fig:theta}
\end{figure}

Via multiplying by a suitable product of tLIOMs, ${\tilde W}_\gamma$ and hence $\theta_\gamma$ can be deformed into ${\tilde W}_{\gamma'}$ and $\theta_{\gamma'}$ with support localized around a path $\gamma'$ (cf. Fig.~\ref{fig:theta}) such that the supports of  $\theta_\gamma$ and $\theta_\gamma'$ do not overlap. 
[In the expansion of $\theta_{\gamma}$ ($\theta_{\gamma'}$) in terms of Pauli strings, the coefficients of operators whose support overlaps with $\gamma'$ ($\gamma$) is exponentially suppressed in the distance between $\gamma$ and $\gamma'$ and hence can be made zero in the $|L|\rightarrow \infty$ limit.]

Since $T_P$ are tLIOMs, we have $[U_F, T_P] = 0$. Together with $[{\tilde W}_\gamma,T_P] = 0$ and $T_P^2=\mathbb{1}$, this implies that $\theta_{\gamma'}=\theta_{\gamma}$ under this deformation despite $\gamma$ and $\gamma'$ being far apart. 
Hence, up to corrections exponentially small in $|L|/\xi$, the operator $\theta_\gamma\propto\mathbb{1}$.
This, together with 
$ \theta_\gamma {\tilde W}_\gamma =  U_F {\tilde W}_\gamma U_F^\dagger$ and ${\tilde W}_\gamma^2=\mathbb{1}$ implies $\theta_\gamma^2 = ( \theta_\gamma \tilde W_\gamma)^2 = \mathbb{1}$, and thus $\theta_\gamma = \pm \mathbb{1}$ to the same accuracy. 

Due to MBL, $U_F$, ${\tilde W}_\gamma$, and hence $\theta_\gamma$ change continuously under perturbation, thus perturbations cannot change the value $\theta_\gamma=-\mathbb{1}$ to $\theta_\gamma=\mathbb{1}$ and vice versa: $\theta_\gamma$ is topologically protected. 
Therefore, if $\{{\tilde W}_\gamma,U_F\} = 0$, this is also protected. The factorization $U_F = \tilde{\m{O}}_L e^{-i f(\{ T_P\},\tilde{W}_\gamma)}$ is thus robust.

What remains to show is that the system is a TTC. This follows from the factorization: it incorporates eigenstate TO by construction and implies $\tilde W_\gamma(mT)=U_F^m \tilde W_\gamma(mT) U_F^{\dagger m}=(-1)^m \tilde W_\gamma(0)$, i.e., period-$2T$ oscillations for the expectation values in Definition~1. These features are robust, owing to the robustness of the factorization. $\qed$

\section{TTCs and absolute stability}
\label{sec:TTCabsst}
The finding of $\tilde W_\gamma(mT) = (-1)^m \tilde W_\gamma(0)$ is an indication that the spectral $\pi$-pairing in $U_F$ is exact for $|L|\rightarrow \infty$, even if $|\gamma|$ is finite. To see this exact $\pi$-pairing explicitly, note that $U_F = \tilde{\m{O}}_L e^{-i f(\{ T_P\},\tilde{W}_\gamma)}$ can be written using Eq.~\eqref{eq:tLIOMs} as
\begin{equation}
U_F = \tilde{\mathcal{O}}'_L e^{-i f_\infty (\{T_P\})},
\end{equation}
where $f_\infty (\{T_P\}) = c_0 + \sum_P c_P T_P + \sum_{P,Q} c_{PQ} T_P T_Q + \ldots$ is the first term in Eq.~\eqref{eq:tLIOMs}, and, by $\{\tilde{\m{O}}_L, {\tilde W}_{\gamma}\} = 0$, 
\begin{align}\label{eq:Oprime}
\tilde{\mathcal{O}}'_L &= e^{i f_{\tilde{W}}(\{T_P\})} \tilde{\mathcal{O}}_L e^{-i f_{\tilde{W}}(\{T_P\})}, \notag \\
 f_{\tilde{W}}(\{T_P\}) &= \frac{\tilde{W}_\gamma}{2} (c^\prime_0 + \sum_P c^\prime_P T_P + \sum_{P,Q} c^\prime_{PQ} T_P T_Q + \ldots ).
\end{align}
The operator $\tilde{\mathcal{O}}'_L$ is a unitary transform of the (dressed) logical Pauli operator $\tilde{\m{O}}_L$. 
It satisfies $[\tilde{\mathcal{O}}'_L,T_P]=0$, hence  %
$[\tilde{\mathcal{O}}'_L,U_F]=0$, so unlike $\tilde{\mathcal{O}}_L$, the operator $\tilde{\mathcal{O}}'_L$ is an integral of motion. 
(Also, since $\{\tilde{\mathcal{O}}'_L,\tilde{W}_\gamma\}=0$, it is a valid conjugate logical operator to $\tilde{W}_\gamma$, albeit not localized near $L$ because  $\tilde{W}_\gamma$ makes $\exp [i f_{\tilde{W}}(\{T_P\})]$ nonlocal.)

Since $\tilde{\mathcal{O}}'_L$ has eigenvalues $\pm 1$, the spectrum of $U_F$ corresponds that of $\pm \exp[-i f_\infty (\{T_P\})]$, implying exact $\pi$-pairing.  
For finite $L$, the $\pi$-pairing receives corrections that decay exponentially in $|L|/\xi$, as follows from the appearance of $\tilde{\m{O}}_L$ in $f$ with such exponentially decaying coefficients. (Such $\tilde{\m{O}}_L$ terms in $f$ when $|L|$ is finite, are present and give exponentially-in-$|L|$ decaying corrections to the $\pi$-pairing even if $|\gamma|\rightarrow \infty$.)

Regular TCs are known to possess ``absolute stability'', i.e., a robustness of $\pi$-spectral pairing or of period doubling even in the presence of symmetry breaking perturbations, with the corrections to these being exponentially suppressed in system size~\cite{KeyserlingkKhemaniSondhi2016stability}. 
The robustness of topological $\pi$-pairing for finite $|\gamma|$ can be seen as a TTC form of this absolute stability. 
In the 1-form symmetry language, the symmetry to consider is $\tilde{\m{O}}_L$. For infinite $|\gamma|$, the Floquet operator commutes with $\tilde{\m{O}}_L$, i.e., the symmetry is present. For finite $|\gamma|$, however, the appearance of ${\tilde W}_\gamma$ (a ``charged object'' under the 1-form symmetry) in $U_F$ spoils this commutation: it is a symmetry breaking perturbation. 
Yet, the $\pi$-pairing is robust so long as $|L|\gg \xi$ (and exact for $|L|/\xi\rightarrow \infty$).

Analogously to regular TCs, this can be traced to the emergence of $\tilde{\mathcal{O}}'_L$ as a new symmetry. 
A key difference from regular TCs, however, is that the symmetry $\tilde{\m{O}}_L$ this replaces was already emergent; it is $\tilde{\m{O}}_L$, replacing the bare $\m{O}_L$ 1-form symmetry (which is generically broken by local perturbations), that is directly analogous to the emergent, local unitary dressed, symmetries in regular TCs. 
The appearance of $\tilde{\mathcal{O}}'_L$ is thus a second layer of symmetry emergence.

\section{Demonstration of Proposition 2}\label{sec:prop2}

For simplicity, we consider a system with a single conjugate pair $\tilde \calO_L,\tilde W_{\gamma}$ of dressed logical operators and, as above, allow finite $|\gamma|$ but take $|L|$ far exceeding the localization length $\xi$ so that $f=f(\{T_P\}, \tilde W_{\gamma})$. 
We inspect
\begin{align}\label{eq:C_alpha}
	&C_\alpha(mT;\gamma,\gamma^\prime) = \langle \alpha | W_\gamma(mT) W_{\gamma^\prime}(0)|\alpha \rangle \notag \\
	&= \sum_{\beta} e^{-i m T(\epsilon_\alpha - \epsilon_\beta)} \langle \alpha | W_\gamma |\beta \rangle \langle \beta | W_{\gamma^\prime} | \alpha \rangle,
\end{align}
where $\gamma^\prime$ is a deformation of $\gamma$ as before and $|\alpha\rangle$, $|\beta\rangle$ are eigenstates of $U_F$ with eigenvalues $e^{-i\epsilon_{\alpha,\beta}T}$. 
In regular TCs one considers long-range correlations~\cite{KeyserlingkKhemaniSondhi2016stability,Khemani2017}; here, we take $\gamma$ and $\gamma^\prime$ to be much further apart from each other than $\xi$. 

Since it is the dressed operators $\tilde{W}_{\gamma^{(\prime)}}$ that show period doubling, we first consider the expansion of $W_{\gamma^{(\prime)}}$ in terms of dressed operators, i.e, products $\tilde{O}$ of operators from the complete set $\{\{T_P, T_P^x\},\tilde W_{\gamma}, \tilde \calO_L\}$  (here $T_P^{x}$ flips $T_P$). 
Afterwards we will assess the time dependence by studying $\tilde{O}(mT) = U_F^m \tilde{O} U_F^{\dagger m}$ for the contributing products $\tilde{O}$.

The expansion coefficients are  $\propto\text{Tr}\left( {W}_{\gamma^{(\prime)}}\tilde{{O}}\right)$. 
Using that $\tilde{{O}} = \tilde U^\dagger {{O}}\tilde U$ (where ${O}$ is the Pauli string corresponding to the dressed operator $\tilde{{O}}$) we find that for a nonzero expansion coefficient, the expansion of $\tilde U W_{\gamma^{(\prime)}} \tilde U^\dagger$ in terms of Pauli strings must include ${O}$, in particular, the support of $\tilde U W_{\gamma^{(\prime)}} \tilde U^\dagger$ must include the entirety of the support of ${O}$. 
Since $\tilde U$ is a local unitary, the support of $\tilde U W_{\gamma^{(\prime)}} \tilde U^\dagger$ is in a ribbon of width $\sim\xi$ around $\gamma^{(\prime)}$, similar to the support of  $\tilde{W}_{\gamma^{(\prime)}}$.
[As in Appendix~\ref{sec:prop1}, we work to exponential accuracy, i.e., ignore the support's tails transversal to $\gamma^{(\prime)}$.]
Therefore, the $\tilde{{O}}$ that can contribute to $W_{\gamma^{(\prime)}}$ must have ${O}$ supported within this ribbon. 

In particular, no product involving $\tilde{\mathcal{O}}_L$ contributes to $W_{\gamma^{(\prime)}}$ for $|L|/\xi\rightarrow \infty$, since $L$ runs off transversally from $\gamma^{(\prime)}$ far beyond the ribbon. (The mutually transversal nature of $\gamma$ and $L$ is due to the Wilson loop algebra $\{W_{\gamma^{(\prime)}},{\mathcal{O}}_L\}=0$ requiring there to be an odd number of intersections between $\gamma$ and $L$, as follows from topological order~\cite{Kitaev2006}.)

Since we took the distance $d(\gamma,\gamma')$ between $\gamma$ and $\gamma'$ to be much larger than $\xi$, products involving $T_P^{x}$ also cannot contribute, as can be seen via the spectral form in Eq.~\eqref{eq:C_alpha}: such a product $\tilde{O}_\gamma^x$ contributing to $W_{\gamma}$ flips tLIOMs near $\gamma$, while $\tilde{O}_{\gamma^{\prime}}^x$ contributing to $W_{\gamma^\prime}$ flips tLIOMs near $\gamma^\prime$. Hence, $\langle \alpha | \tilde{O}_\gamma^x |\beta \rangle \langle \beta | W_{\gamma^\prime} | \alpha \rangle=\langle \alpha | W_\gamma |\beta \rangle \langle \beta | \tilde{O}_{\gamma^{\prime}}^x | \alpha \rangle = 0$ [up to exponentially small corrections in $d(\gamma,\gamma')/\xi$] for all eigenstates $|\alpha\rangle$, $|\beta\rangle$.

The only surviving contributions are those from $\{T_P \}$ and $\tilde W_{\gamma}$. 
Since for any product $\tilde{O}_0=\prod_P T_{P}$ and $\tilde{O}_1= \tilde W_{\gamma} \tilde{O}_0$ we have $U_F^m \tilde{O}_j U_F^{m\dagger}=(-1)^{jm} \tilde{O}_j$, we find using Eq.~\eqref{eq:C_alpha} 
\begin{align}
	C_\alpha(mT;\gamma,\gamma^\prime) = c_0(\gamma,\gamma^\prime;\alpha) + c_1(\gamma, \gamma^\prime;\alpha) (-1)^m, \label{eq:splitting}
\end{align}
where the $c_j$ terms come from $\tilde{O}_j$-type contributions to $W_\gamma$. 
The TTC signal strength is controlled by $c_1$. 
In terms of the sum $\tilde{\mathcal{S}}_{j\gamma^{(\prime)}}$ of all $\tilde{O}_{j}$-type contributions  to $W_{\gamma^{(\prime)}}$, we have
$c_{1}(\gamma,\gamma^{\prime};\alpha)=\langle\alpha|\tilde{\mathcal{S}}_{1\gamma}(\tilde{\mathcal{S}}_{0\gamma^\prime}+\tilde{\mathcal{S}}_{1\gamma^\prime})|\alpha\rangle$.

We next upper bound $|c_{1}(\gamma,\gamma^{\prime};\alpha)|$. 
By the Cauchy-Schwartz inequality, $|c_{1}(\gamma,\gamma^{\prime};\alpha)|^2\leq \langle\alpha|\tilde{\mathcal{S}}_{1\gamma}^2 |\alpha\rangle \langle\alpha|(\tilde{\mathcal{S}}_{0\gamma^\prime}+\tilde{\mathcal{S}}_{1\gamma^\prime})^2 |\alpha\rangle$. 
We focus on $\langle\alpha|\tilde{\mathcal{S}}_{1\gamma}^2 |\alpha\rangle$ for brevity; analogous considerations hold for $\langle\alpha|(\tilde{\mathcal{S}}_{0\gamma^\prime}+\tilde{\mathcal{S}}_{1\gamma^\prime})^2 |\alpha\rangle$. 
Due to $\langle\alpha|\tilde{\mathcal{S}}_{1\gamma}^2 |\alpha\rangle\geq 0$, and due to MBL no eigenstate being special, one may estimate 
\begin{equation}
\langle\alpha|\tilde{\mathcal{S}}_{1\gamma}^2 |\alpha\rangle\sim \frac{1}{2^N}\sum_\alpha \langle\alpha|\tilde{\mathcal{S}}_{1\gamma}^2 |\alpha\rangle = \frac{1}{2^N}\Tr[\tilde{\mathcal{S}}_{1\gamma}^2],
\end{equation}
where $N$ is the number of qubits in the system. 
Furthermore,
$\text{Tr}[\tilde{\mathcal{S}}_{1\gamma}^{2}]/2^{N}=\sum_{j\in\tilde{\mathcal{S}}_{1\gamma}}|\eta_{j\gamma}|^{2}$ in terms of the coefficients $\eta_{j\gamma}$ from the expansion of $W_{\gamma}$ in terms of $\{T_{P},T_{P}^{x}\}$ and $\tilde{W}_{\gamma}$.

As noted above, only such $\tilde{O}_{1}$ contribute to $\tilde{\mathcal{S}}_{1\gamma}$ that have support within a ribbon of width $\sim\xi$ around $\gamma$. 
The number of such terms in $\tilde{\mathcal{S}}_{1\gamma}$ equals the number of different products the $\sim|\gamma|\xi$ tLIOMs $T_P$ in this ribbon can result in: this is $\sim2^{|\gamma|\xi}$ terms. 
Within the same ribbon, the expansion of $W_\gamma$ however also includes $T_P^x$ and $T_P T_P^x$ factors, hence it has $\sim 4^{|\gamma|\xi}$ terms. 
Due to MBL, the expansion has no structure within the ribbon, hence, $|\eta_{j\gamma}|^2\sim4^{-|\gamma|\xi}$ due to normalization. 
Hence, $\text{Tr}[\tilde{\mathcal{S}}_{1\gamma}^{2}]/2^{N}\lesssim 2^{|\gamma|\xi} 4^{-|\gamma|\xi} =  2^{-|\gamma|\xi}$. 
The same logic holds for $\langle\alpha|(\tilde{\mathcal{S}}_{0\gamma^\prime}+\tilde{\mathcal{S}}_{1\gamma^\prime})^2 |\alpha\rangle$, hence
$|c_{1}(\gamma,\gamma^{\prime};\alpha)|\lesssim2^{-(|\gamma|+|\gamma'|)\xi/2}$.
For finite $|\gamma^{(\prime)}|$ the coefficient $c_1$ is nonetheless nonzero.

\section{Comparison of regular and topological time crystals}\label{sec:TCvsTTC}

\begin{figure}[h!]
\includegraphics[width=0.3\textwidth]{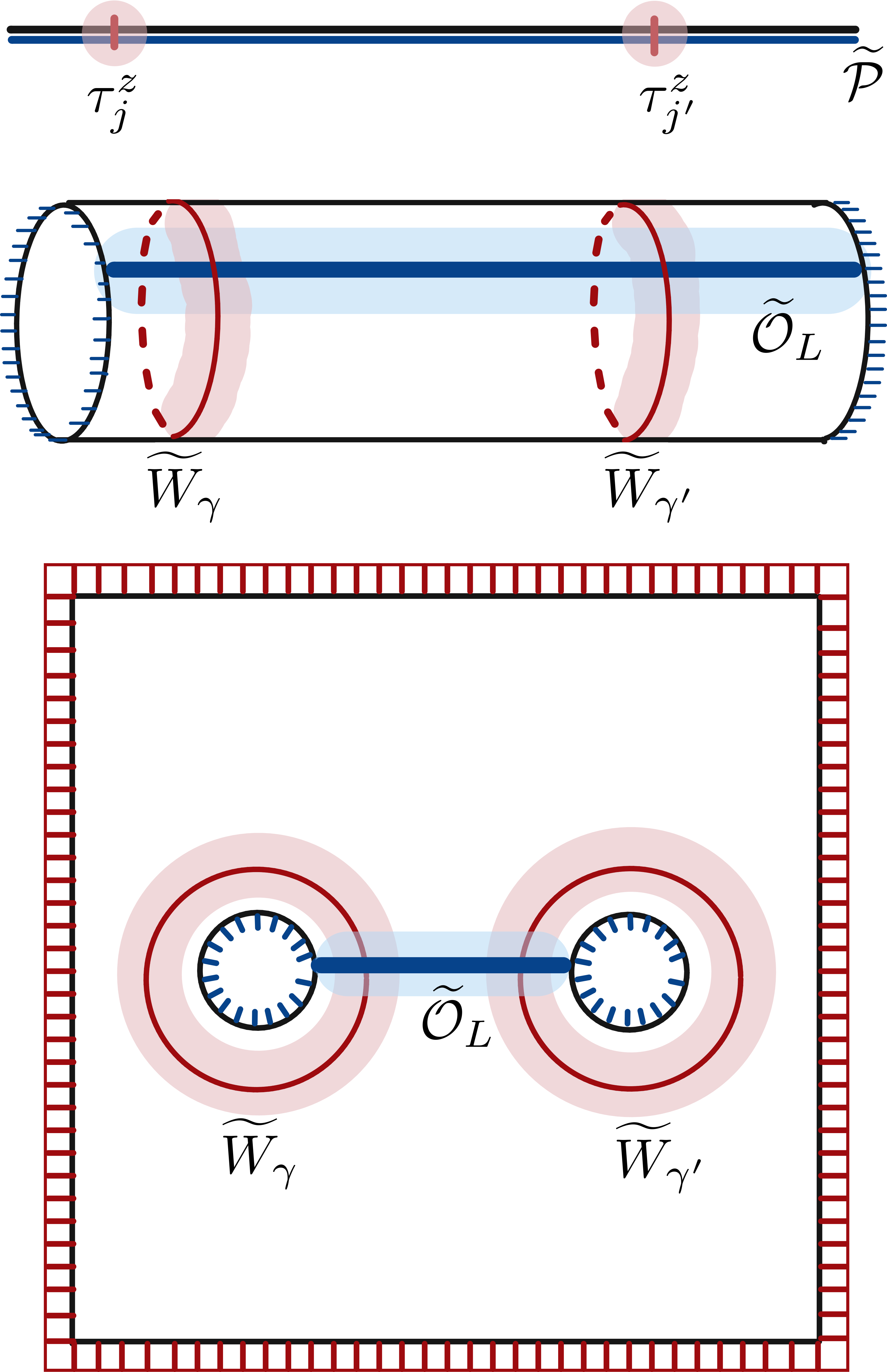}
\caption{A visual comparison of regular TCs and TTCs. Top panel: a regular TC on a 1D lattice (black line). The emergent symmetry $\tilde{\mathcal{P}}$ (blue line) acts at every site. Red tickmarks indicate Pauli operators $Z_{j^{(\prime)}}$, while opaque disks, reminding of the exponential localization at scale $\sim \xi$, indicate $\tau_{j^{(\prime)}}^z$.
Middle panel: 
TTC on a cylinder. The blue and red lines show, respectively, the bare logical operator ${\mathcal{O}}_L$ and $W_{\gamma^{(\prime)}}$, while their smeared opaque counterparts, reminding of the exponential localization at scale $\sim \xi$ transverse to the operator path, stand for $\tilde{\mathcal{O}}_L$ and $\tilde{W}_{\gamma^{(\prime)}}$. The decoration at the edges indicates  boundary conditions such that ${\mathcal{O}}_L$ is along the cylinder.
Bottom panel: planar geometry topologically equivalent to the middle panel's cylinder. The boundary decorations indicate boundary conditions such that ${\mathcal{O}}_L$ runs between the holes, while $W_\gamma$ runs around either of the holes. [$W_\gamma$ and $W_{\gamma'}$ encircling different holes can be obtained from one another via multiplication by bare stabilizers (not shown).]
}
\label{fig:TCvsTTC}
\end{figure}

As noted in the main text, the TTCs we consider, constructed by lifting the global symmetry considerations of regular TCs to higher-form symmetries, give arguably the most ``natural'' forms of TCs with intrinsic TO. 
Here we provide further details on the respective regular TC and TTC aspects by tabulating the various ingredients we used in constructing TTCs and indicating their regular TC counterparts.
(As in the main text, we focus on qubit systems displaying $\mathbb{Z}_2$ TO.) 
We summarize these ingredients in Table~\ref{tab:TCvsTTC} and illustrate them in
Fig.~\ref{fig:TCvsTTC}. 

\renewcommand{\arraystretch}{1.5}

\begin{table*}
\begin{centering}
\begin{tabular}{>{\centering}m{6cm}|>{\centering}m{6cm}|>{\centering}m{6cm}}
Ingredient & Regular TC & Topological TC\tabularnewline
\hline 
\hline 
Microscopic global symmetry & $\mathbb{Z}_{2}$ symmetry, e.g., $\mathcal{P}=\prod_{j}X_{j}$ & $\mathbb{Z}_{2}$ 1-form symmetry (bare logical operator) $\mathcal{O}_{L}$,
e.g., $\mathcal{O}_{L}=\prod_{j\in L}X_{j}$\tabularnewline
\hline 
Stabilizer code  & Repetition code with stabilizers $\{Z_{j}Z_{j+1}\}$ and logical operators $\mathcal{P}$, $Z_{j}$  & Topological code with stabilizers $\{A_{P}\}$ and logical operators
$\mathcal{O}_{L}$, $W_{\gamma}$ \tabularnewline
\hline 
Emergent symmetry & $\tilde{\mathcal{P}}=\tilde{U}^{\dagger}\mathcal{P}\tilde{U}$ & $\tilde{\mathcal{O}}_{L}=\tilde{U}^{\dagger}\mathcal{O}_{L}\tilde{U}$
[or $\tilde{\mathcal{O}}_{L}^{\prime}$ in Eq.~\eqref{eq:Oprime}]\tabularnewline
\hline 
Local integrals of motion  & $\tilde{D}_{j}=\tilde{U}^{\dagger}Z_{j}Z_{j+1}\tilde{U}$ & $T_{P}=\tilde{U}^{\dagger}A_{P}\tilde{U}$\tabularnewline
\hline 
Symmetry-odd $\mathcal{M}$ satisfying $\mathcal{M}(mT)=(-1)^{m}\mathcal{M}(0)$ & $\tau_{j}^{z}=\tilde{U}^{\dagger}Z_{j}\tilde{U}$  & $\tilde{W}_{\gamma}=\tilde{U}^{\dagger}W_{\gamma}\tilde{U}$  \tabularnewline
\hline 
Correlator for time crystal signatures in eigenstates $|\alpha\rangle$ & $\langle\alpha|Z_{j}(mT)Z_{j'}(0)|\alpha\rangle,\qquad|j-j'|\gg\xi$ & $\langle\alpha|W_{\gamma}(mT)W_{\gamma'}(0)|\alpha\rangle,\qquad d(\gamma,\gamma^{\prime})\gg\xi$\tabularnewline
\hline 
\end{tabular}
\par\end{centering}
\caption{Comparing the ingredients of regular TCs and TTCs. A detailed description of the entries is provided in the text, while a graphical illustration emphasizing the locality aspects is given in Fig.~\ref{fig:TCvsTTC}.
}\label{tab:TCvsTTC}
\end{table*}

We start with regular TCs. 
A brief summary of ingredients is as follows~\cite{KeyserlingkKhemaniSondhi2016stability,Khemani2017}.
The microscopic global symmetry is $\mathcal{P}=\prod_{j}X_{j}$ where the subscript $j$ labels the sites of the system. 
The local interactions respecting this symmetry include the bare stabilizers $D_{j}=Z_{j}Z_{j+1}$ of the repetition code.
This code has logical operators $\mathcal{P}$ and $Z_{j}$; note that  $Z_j$ and $Z_{j'}$ are equivalent logical operators, obtainable from one another via multiplication by the stabilizers $D_j$. 
Absolute stability means that the microscopic symmetry need not be preserved by the system; 
in the presence of weak perturbations, it is replaced by an emergent symmetry $\tilde{\mathcal{P}}=\prod_{j}\tau_{j}^{x}$ where $\tau_{j}^{x}=\tilde{U}^{\dagger}X_{j}\tilde{U}$ are the local unitary dressed (and hence exponentially localized on the scale of $\xi$) counterparts of $X_{j}$. 
The corresponding local integrals of motion are $\tilde{D}_{j}=\tau_{j}^{z}\tau_{j+1}^{z}$. %
The TC Floquet unitary has the form $U_{F}=\tilde{\mathcal{P}}e^{-if(\{\tilde{D}_{j}\})}$
with an exponentially local function $f$. 
Period doubling is present in any operator that commutes with all $\tilde{D}_{j}$
and anticommutes with the symmetry $\tilde{\mathcal{P}}$. 
For $\tau_{j}^{z}=\tilde{U}^{\dagger}Z_{j}\tilde{U}$, in particular, we have $[\tau_{j}^{z},\tilde{D}_{j}]=\{\tau_{j}^{z},\tilde{\mathcal{P}}\}=0$
and thus $\tau_{j}^{z}(mT)=(-1)^{m}\tau_{j}^{z}(0)$. 
Since the concrete form of $\tau_{j}^{z}$ depends
on the disorder realization (due to this setting $\tilde{U}$), to detect TC behavior it is better to use the bare logical operators $Z_{j}$: 
owing to $\tau_{j}^{z}$'s exponential localization, the $Z_j$ are good approximants of $\tau_{j}^{z}$ %
and their correlations $\langle\alpha|Z_{j}(mT)Z_{j'}(0)|\alpha\rangle$ in eigenstates $|\alpha\rangle$ reveal period doubling, provided $|j-j'|\gg\xi$ so that spurious effects from the overlaps of exponential tails (in the expansion of $Z_{j}$ in terms of $\tau_{i}^{\alpha}$) are eliminated.

We now turn to TTCs. 
The parallels between regular TCs and TTCs, especially in terms of locality, are the most apparent from comparing the TTC in the middle panel of Fig.~\ref{fig:TCvsTTC} with the TC in the top panel.
The summary of the TTC construction, in a similar order to that for TCs above, and focusing on this cylinder system is as follows. 
The two ends of the cylinder have such boundary conditions that the system supports the logical operator $\mathcal{O}_{L}$ along the cylinder; the conjugate logical operator $W_{\gamma}$ runs around the cylinder~\cite{dennisTopologicalQuantumMemory2002,FowlerMariantoniMartinisCleland2012surfacecode,TerhalRMP}.
(Equivalent logical operators differ from one another by deformations of their path; if $W_{\gamma}$ and $W_{\gamma'}$ are two such equivalent logical operators, they are obtainable from one another via multiplication by topological stabilizers $A_{P}$.)
Now the microscopic 1-form symmetry is $\mathcal{O}_{L}$ (for concreteness, one can imagine this as $\mathcal{O}_{L}=\prod_{j\in L}X_{j}$) and local interactions that respect (any deformation of) $\mathcal{O}_{L}$ include the bare topological stabilizers $A_P$. 
Under weak, symmetry-breaking, perturbations, we have the emergent symmetry $\tilde{\mathcal{O}}_{L}=\tilde{U}^{\dagger}\mathcal{O}_{L}\tilde{U}$ with the local unitary $\tilde{U}$. 
[More precisely, the emergent symmetry is Eq.~\eqref{eq:Oprime} for a cylinder of finite circumference.] 
The local integrals of motion are the tLIOMs $T_{P}=\tilde{U}^{\dagger}A_{P}\tilde{U}$.
The Floquet unitary has the form $U_{F}=\tilde{\mathcal{O}}_{L}e^{-if(\{T_{P},\tilde{W}_{\gamma}\})}$ with an exponentially local function $f$ and $\tilde{W}_{\gamma}=\tilde{U}^{\dagger}W_{\gamma}\tilde{U}$.
(See Sec.~\ref{sec:prop1} above for a detailed description of $f$ for a cylinder of finite circumference.)
Period doubling is present in any operator that commutes with all $T_{P}$ and $\tilde{W}_{\gamma}$ and anticommutes with $\tilde{\mathcal{O}}_{L}$. 
In particular we have $[\tilde{W}_{\gamma},T_{P}]=[\tilde{W}_{\gamma},\tilde{W}_{\gamma'}]=\{\tilde{W}_{\gamma},\tilde{\mathcal{O}}_{L}\}=0$ and thus $\tilde{W}_{\gamma}(mT)=(-1)^{m}\tilde{W}_{\gamma}(0)$. 
Since the concrete form of $\tilde{W}_{\gamma}$
depends on the disorder realization via $\tilde{U}$, to detect TTC behavior it is better to use the bare $W_{\gamma}$:
these are concretely given Pauli strings that, owing to $\tilde{W}_{\gamma}$'s exponential localization transverse to its path, are good approximants of $\tilde{W}_{\gamma}$ 
and their correlations
$\langle\alpha|W_{\gamma}(mT)W_{\gamma'}(0)|\alpha\rangle$ in eigenstates $|\alpha\rangle$ reveal period doubling, provided the distance $d(\gamma,\gamma')$ between the paths $\gamma$ and $\gamma'$ satisfies $d(\gamma,\gamma')\gg\xi$ so that spurious effects from the overlaps of exponential tails (in the expansion of $W_{\gamma}$ in terms of the dressed operators, cf. the demonstration of Proposition 2) are eliminated.

As Proposition 2 shows, a subtlety in the topological case is that for the signal to survive the thermodynamic limit, one must keep  the path length $|\gamma|$ finite. 
For the cylinder in the middle panel of Fig.~\ref{fig:TCvsTTC}, this amounts to a thermodynamic limit with fixed cylinder circumference.
While the resulting system is quasi-1D, we emphasize that working with quasi-1D systems in not a requirement for TTCs. 
This is illustrated in the bottom panel of Fig.~\ref{fig:TCvsTTC}, which shows a system topologically equivalent to the cylinder: a planar system with two holes on which $\mathcal{O}_{L}$ terminates.
(The outer boundary, conversely, allows the termination of $W_{\gamma}$.) 
In this case, the thermodynamic limit can result in a 2D system; the requirement is now that the hole circumferences remain finite. 
Long-range correlations correspond to correlations of $W_{\gamma}$ encircling distinct, far-separated holes in the system.

\end{document}